\documentclass[aps,10pt,prd,eqsecnum,twocolumn,nofootinbib]{revtex4-1}

\usepackage{amsfonts}
\usepackage{amsmath}
\usepackage{amssymb}
\usepackage{amsthm}
\usepackage{bm}
\usepackage{dcolumn}
\usepackage{epsfig}
\usepackage{graphicx}
\usepackage{graphics}
\usepackage[latin1]{inputenc}
\usepackage{latexsym}
\usepackage{rotating}

\begin{document}

\title{Bound orbits of a slowly evolving black hole}

\author{Scott A.\ Hughes}
\affiliation{Department of Physics and MIT Kavli Institute, Cambridge, MA 02139}

\begin{abstract}
Bound orbits of black holes are very well understood.  Given a Kerr black hole of mass $M$ and spin $S = aM^2$, it is simple to characterize its orbits as functions of the orbit's geometry.  How do the orbits change if the black hole is itself evolving?  How do the orbits change if the orbiting body evolves? In this paper, we consider a process that changes a black hole's mass and spin, acting such that the spacetime is described by the Kerr solution at any moment, or that changes the orbiting body's mass.  Provided this change happens slowly, the orbit's actions ($J_r, J_\theta, J_\phi$) are {\it adiabatic invariants}, and thus are constant during this process.  By enforcing adiabatic invariance of the actions, we deduce how an orbit evolves due to changes in the black hole's mass and spin and in the orbiting body's mass.  We demonstrate the impact of these results with several examples: how an orbit responds if accretion changes a black hole's mass and spin; how it responds if the orbiting body's mass changes due to accretion; and how the inspiral of a small body into a black hole is affected by change to the hole's mass and spin due to the gravitational radiation absorbed by the event horizon.  In all cases, the effect is very small, but can be an order of magnitude or more larger than what was found in previous work which did not take into account how the orbit responds due to these effects.
\end{abstract}

\maketitle

\section{Introduction and motivation}

Computing orbits in a static gravitational potential $V({\bf r})$ (in Newtonian theory) or in a stationary spacetime $g_{\mu\nu}({\bf x})$ (in general relativity) is a standard and important problem in mechanics.  How do these orbits change if the potential or the spacetime changes with time?

A clear answer exists in a particular limit for the Newtonian version of this problem.  Imagine a gravitational potential $V({\bf r};t)$ that continually and smoothly evolves over a time interval $t_i \le t \le t_f$.  Suppose that the potential's orbits are integrable at each moment in this interval, and suppose further that the evolution is ``slow,'' in the sense that the potential's change is very small over a single orbital period $T_{\rm orb}$:
\begin{equation}
\biggl|\frac{T_{\rm orb}}{V}\frac{\partial V}{\partial t}\biggr| \ll 1\;.
\label{eq:slowevolve}
\end{equation}
Since motion in this potential is integrable, one can identify a set of canonical coordinates $x^k$ and their conjugate momenta $p_k$ that serve as particularly useful labels for the system in phase space.  In particular, such orbits live on the surface of a 6-torus, each of which is labeled by a set of three {\it action variables}, or actions
\begin{equation}
J_k = \frac{1}{2\pi}\oint p_k\,dx^k\;,
\label{eq:Jkdef}
\end{equation}
where $k$ labels the three canonical coordinates that are used in this construction.  These actions quantify the phase-space area enclosed by the orbit.  It can then be shown these actions are {\it adiabatic invariants}, and as such remain fixed in value while the potential changes: $J_k \to J_k$ as $V \to V + \delta V$, provided Eq.\ (\ref{eq:slowevolve}) is satisfied.

A proof of adiabatic invariance for actions can be found in many dynamics textbooks; particularly concise and clear discussion is given in Binney and Tremaine (Ref.\ {\cite{bt1987}}, Sec.\ 3.6).  This proof of adiabatic invariance has three critical ingredients: The motion must be integrable, so that all orbits are characterized by actions (\ref{eq:Jkdef}); the system must evolve from one integrable configuration to another, so that actions exist for every configuration as it evolves; and the evolution must be slow, in the sense that a suitable generalization of Eq.\ (\ref{eq:slowevolve}) describes how the system evolves.  As long as these three conditions are met, the actions describing a given system are adiabatic invariants.  A version of this proof, which borrows heavily from Ref.\ {\cite{bt1987}}, is given in Appendix {\ref{app:proof}}.

Geodesic orbits of Kerr black holes are fully integrable {\cite{carter1968}}, and it is well known that they can be characterized by a set of actions associated with their three spatial coordinate motions.  If we imagine a Kerr spacetime which is slowly evolving, but doing so in such a way that it evolves from one Kerr spacetime to another, then the proof of adiabatic invariance applies to the actions of Kerr black hole orbits.  In other words, provided the spacetime is the Kerr solution at each moment, then $J_k \to J_k$ as the black hole's mass and spin evolve according to $M \to M + \delta M$, $S \to S + \delta S$, or as the orbiting body's mass changes according to $\mu \to \mu + \delta\mu$.  (It should be emphasized that adiabatic invariance does not hold for all adiabatic evolution mechanisms.  Dissipative evolution mechanisms in particular, such as the backreaction of gravitational-wave emission, change $J_k$.  As such, it is important in analyzing the mechanisms which can evolve a system to distinguish and separate those terms which can change the actions from those for which adiabatic invariance will hold.)

The remainder of this paper examines the consequences of this invariance.  We begin in Sec.\ {\ref{sec:newton}} with a worked example in Newtonian gravity.  This allows us to illustrate this concept in a simple limit.  We then turn to orbits of Kerr black holes, first examining the simple case of equatorial circular orbits in Sec.\ {\ref{sec:kerr_circeq}}.  For both Newtonian gravity and circular equatorial Kerr black hole orbits, we also compute the (incorrect) result one would obtain evolving $M$, $S$, or $\mu$ without enforcing adiabatic invariance, but instead assuming that the orbit's parameters $(p,e,I)$ are fixed during this evolution (an assumption that has been used in past work which examined binaries around slowly evolving black holes).  We show that the incorrect variation typically underestimates how the system responds to this evolution, in some cases by a rather large factor.

To develop some intuition for the generic case, we next examine the weak-field expansion of equatorial and eccentric orbits in Sec.\ {\ref{sec:eq_wf}}, and then finally go to the fully generic case in Sec.\ {\ref{sec:kerr_generic}}.  We cannot present closed-form results for the generic case, but we describe the calculation in detail, and give numerical examples illustrating how orbits change due to evolution of either the black hole's mass and spin or of the mass of the orbiting body.

In Sec.\ {\ref{sec:inspiral}}, we examine how correctly accounting for adiabatic invariance affects models for the inspiral of a small body into a Kerr black hole.  We examine two problems that have been discussed in past literature: a black hole whose mass and spin changes during the inspiral due to accretion (discussed in, for example, Ref.\ {\cite{bcp14}}); and a black hole whose mass and spin changes during the inspiral due to the fraction of gravitational radiation that is absorbed by the hole's event horizon (discussed recently in Ref.\ {\cite{in2018}}).  We compute how much orbital phase shift is introduced by these effects as compared to a model that leaves the black hole mass and spin unchanged.  Previous work found that these effects were virtually negligible.  Our analysis shows that correctly accounting for adiabatic invariance of the actions typically increases the phase shift by about an order of magnitude (less in some cases, more in others).

Since ten times a nearly infinitesimal phase shift remains a nearly infinitesimal phase shift, we do not disagree with the conclusions of previous work.  Nonetheless, as we summarize in our conclusions (Sec.\ {\ref{sec:conclude}}), it is at minimum salubrious to correct the underlying physics of these methods; and, our results indicate that there may be circumstances where these ``virtually negligible'' effects may not be quite so ignorable.  We also note that the effects we find in our analysis of radiation absorbed by the black hole's event horizon should also be present in a self-consistent self force analysis which includes the backreaction of a body upon the Kerr spacetime.  Given the challenge of computing many self force effects, particularly as the community begins developing methods to go beyond leading order, it may be valuable to have relatively simple-to-compute invariants with which certain kinds of self force results can be compared.

Certain technical details are relegated to appendices.  As already mentioned, Appendix {\ref{app:proof}} gives a brief proof of adiabatic invariance for the actions under the circumstances that we consider, borrowing heavily from Ref.\ {\cite{bt1987}}.  Appendix {\ref{app:actions}} describes how we compute the actions of Kerr black hole orbits, as well as the derivatives of the actions which are needed for many of our computations.  Appendix {\ref{app:flux}} presents the gravitational-wave flux formulae that we use for our analyses in Sec.\ {\ref{sec:inspiral}}.

Throughout this paper, we use units with $G = 1 = c$.

\section{Background: Adiabatic invariance in an evolving Newtonian potential}
\label{sec:newton}

For intuition, we begin with a simple, illustrative case: a test mass $\mu$ orbiting a spherical body $M$ in Newtonian gravity, with $\mu \ll M$.  Take its orbit to have semi-latus rectum $p$ and eccentricity $e$, so that it oscillates radially from periapsis $r_p = p/(1 + e)$ to apoapsis $r_a = p/(1 - e)$.  Let the orbit be inclined to the equatorial plane by an angle $I$ that lies between 0 and $\pi$ radians; the orbit thus oscillates in polar angle from
\begin{equation}
\theta_{\rm min} = \pi/2 - I\;\quad\mbox{to}\quad \theta_{\rm max} = \pi/2 + I
\label{eq:thetaminmax_pro}
\end{equation}
for $I \le \pi/2$, and from
\begin{equation}
\theta_{\rm min} = I - \pi/2\;\quad\mbox{to}\quad\theta_{\rm max} = 3\pi/2 - I\;
\label{eq:thetaminmax_ret}
\end{equation}
for $I > \pi/2$.  This orbit has energy
\begin{equation}
E = -\frac{\mu M}{2p}(1 - e^2)\;,
\end{equation}
it has an angular momentum about the $z$ axis
\begin{equation}
L_z = \mu\sqrt{Mp}\cos I\;,
\end{equation}
and it has an angular momentum normal to the $z$ axis
\begin{equation}
L_\perp = \mu\sqrt{Mp}\sin I\;.
\end{equation}
The orbit's three actions are given by
\begin{eqnarray}
J_\phi &=& \frac{1}{2\pi}\int_0^{2\pi}p_\phi\,d\phi = \mu\sqrt{Mp}\cos I\;,
\label{eq:JphiNewt}\\
J_\theta &=& \frac{1}{\pi}\int_{\theta_{\rm min}}^{\theta_{\rm max}}p_\theta\,d\theta = \mu\sqrt{Mp}\left(1 - \cos I\right)\;,
\label{eq:JthetaNewt}\\
J_r &=& \frac{1}{\pi}\int_{r_p}^{r_a}p_r\,dr = \mu\sqrt{Mp}\left[(1 -e^2)^{-1/2} - 1\right]\;.
\nonumber\\
\label{eq:JrNewt}
\end{eqnarray}
Notice that $J_\phi = L_z$.  This is a generic result, holding for any orbit in axisymmetric potentials or spacetimes.

Now imagine that the mass of the gravitating body increases: $M \to M + \delta M$.  As long as this process is slow in the sense of Eq.\ (\ref{eq:slowevolve}), then the actions $J_{\phi,\theta,r}$ will be adiabatic invariants: we must have $\delta J_{\phi,\theta,r} = 0$ as $M \to M + \delta M$.  Enforcing
\begin{eqnarray}
0 = \delta J_k = \frac{\partial J_k}{\partial M}\delta M + \frac{\partial
  J_k}{\partial p}\delta p + \frac{\partial J_k}{\partial e}\delta e +
\frac{\partial J_k}{\partial I}\delta I
\nonumber\\
\end{eqnarray}
for $k \in [r,\theta,\phi]$, we find
\begin{equation}
\delta p = -p\frac{\delta M}{M}\;,\quad \delta e = 0\;,\quad \delta I = 0\;.
\label{eq:Newton_deltap}
\end{equation}
An increase in the gravitating body's mass causes the orbit to shrink, doing so in a way that leaves its shape (its eccentricity and inclination) unchanged.  It is not surprising that the orbit's inclination does not change, since the variation we consider is spherically symmetric and has no effect on the orbit's angular momentum.  The orbit's energy is changed by this process:
\begin{equation}
\delta E = \frac{\partial E}{\partial M}\delta M + \frac{\partial E}{\partial p}\delta p
+ \frac{\partial E}{\partial e}\delta e = -\frac{\mu}{p}(1 - e^2)\delta M\;.
\end{equation}
This is also not surprising: the system's evolution is not time independent, so we do not expect the process to conserve energy.

The observationally significant effect arising from the change to $M$ will be a change to the orbit's frequency $\Omega$.  This frequency is given by Kepler's third law, which when expressed in terms of $p$ and $e$ takes the form
\begin{equation}
\Omega = \sqrt{\frac{M(1-e^2)^3}{p^3}}\;.
\label{eq:Kepler3}
\end{equation}
Allowing the mass to change and enforcing adiabatic invariance, we find
\begin{equation}
\delta\Omega = \frac{\partial\Omega}{\partial M}\delta M + \frac{\partial\Omega}{\partial p}\delta p
+ \frac{\partial\Omega}{\partial e}\delta e = 2\Omega\frac{\delta M}{M}\;.
\label{eq:NewtonianOmegaChange}
\end{equation}
Note that if we only adjusted the mass and left out the change in the orbit that comes from enforcing adiabatic invariance, we would get a smaller result:
\begin{equation}
\delta\Omega_{\rm wrong} \equiv \frac{\partial\Omega}{\partial M}\delta M = \frac{1}{2}\Omega\frac{\delta M}{M} = \frac{1}{4}\delta\Omega\;.
\end{equation}
This illustrates how the adiabatically invariant response of an orbit to a change in its source can in principle have an observationally important impact.

Adiabatic invariance allows us to examine another process: how the orbit responds due to a change in the mass $\mu$ of the orbiting body.  Such a change could occur if the orbiting body were embedded in an accretion flow, for example.  In a realistic scenario of this form, the accreting matter would exert a torque on the orbit due to, for example, the viscosity of the accreting fluid.  For illustrative purposes, let us imagine that this material only changes $\mu$ and examine how the orbit responds to this change.  Enforcing adiabatic invariance now means
\begin{eqnarray}
0 = \delta J_k = \frac{\partial J_k}{\partial \mu}\delta \mu + \frac{\partial
  J_k}{\partial p}\delta p + \frac{\partial J_k}{\partial e}\delta e +
\frac{\partial J_k}{\partial I}\delta I
\nonumber\\
\end{eqnarray}
for $k \in [r,\theta,\phi]$.  Solving this system, we find
\begin{equation}
\delta p = -2p\frac{\delta\mu}{\mu}\;,\quad \delta e = 0\;,\quad \delta I = 0\;.
\label{eq:Newton_deltap2}
\end{equation}
The small body spirals inward as its mass grows, holding its shape fixed.  This likewise changes the orbital frequency:
\begin{equation}
\delta\Omega = \frac{\partial\Omega}{\partial p}\delta p + \frac{\partial\Omega}{\partial e}\delta e = 3\Omega\frac{\delta\mu}{\mu}\;.
\label{eq:NewtonianOmegaChange2}
\end{equation}
Without correctly accounting for adiabatic invariance, one expects $\delta\Omega_{\rm wrong} = 0$ since the mass of the gravitating source is unchanged --- a dramatically different result from the correct one.

\section{Black hole orbits I:\\ Circular and equatorial}
\label{sec:kerr_circeq}

We now apply adiabatic invariance to the actions that characterize black hole orbits.  We start with the simple case of circular and equatorial orbits: orbits whose radius $r$ is fixed, and which lie in the plane normal to the black hole's spin axis, $\theta = \pi/2$.  Such orbits are amply discussed in Ref.\ {\cite{bpt72}}; their properties are sufficiently simple that we can find exact expressions for all important aspects of this analysis.  We refer the reader to {\cite{bpt72}} for derivation and details of the results we use below.

For circular and equatorial orbits of Kerr black holes, the actions are given by
\begin{eqnarray}
J_\phi &=& L_z = \pm\mu M \frac{1 \mp 2 a v^3 + a^2 v^4}{v\sqrt{1 - 3 v^2 \pm 2 a v^3}}\;,
\label{eq:circeq_Jphi}
\\
J_r &=& J_\theta = 0\;.
\label{eq:circeq_Jr_Jtheta}
\end{eqnarray}
Here, $\mu$ is the mass of the orbiting body, $a = S/M^2$ is the dimensionless spin parameter of the black hole, and $v = \sqrt{M/r}$, where $r$ is orbital radius.  The parameter $v$ is, roughly speaking, the speed of the orbit, and the spin parameter lies in the range $0 \le a \le 1$.  The upper signs in Eq.\ (\ref{eq:circeq_Jphi}) and in all equations which follow describe prograde orbits, with angular momentum parallel to the hole's spin; the lower signs describe retrograde orbits.  These orbits have frequency
\begin{equation}
\Omega = \pm \frac{M^{1/2}}{r^{3/2} \pm a M^{3/2}}
= \pm\frac{v^3}{M(1 \pm av^3)}\;.
\label{eq:circeq_Omega}
\end{equation}
This is the frequency conjugate to the Boyer-Lindquist coordinate time, and thus describes the frequency of the orbit as measured by the clocks of distant observers.

Consider a process that takes $M \to M + \delta M$, $S \to S + \delta S$.  Using
\begin{eqnarray}
\delta v &=& \frac{v}{2}\left(\frac{\delta M}{M} - \frac{\delta r}{r}\right)\;,
\\
\delta a &=& a \left(\frac{\delta S}{aM^2} - 2\frac{\delta M}{M}\right)\;,
\end{eqnarray}
it is straightforward to show that $J_\phi \to J_\phi$ implies
\begin{eqnarray}
\frac{\delta r}{r} &=& -\frac{1 \pm 3av^3(1 - 2v^2) - a^2v^4(3 - 10v^2) \mp 5a^3v^7}{(1 \pm av^3)(1 - 6v^2 \pm 8av^3 - 3 a^2v^4)}\frac{\delta M}{M}
\nonumber\\
& & \pm \frac{6v^3(1 - 2v^2) \mp 4av^4(1 - 4v^2) - 6a^2v^7}{(1 \pm av^3)(1 - 6v^2 \pm 8av^3 - 3 a^2v^4)}\frac{\delta S}{M^2}\;.
\label{eq:circeq_deltar}
\end{eqnarray}
This simplifies significantly in the Schwarzschild limit:
\begin{equation}
\frac{\delta r}{r} \to -\frac{1}{1 - 6v^2}\frac{\delta M}{M}
\pm \frac{6v^3(1 - 2v^2)}{1 - 6v^2}\frac{\delta S}{M^2}
\end{equation}
as $a \to 0$.  The term in $\delta M$ is consistent with the Newtonian result (\ref{eq:Newton_deltap}), although note the singularity at the innermost stable circular orbit, $r = 6M$.  This reflects the fact that in the strong field of a black hole, a slight change to the spacetime may have a large effect on an orbit that is close to the last stable orbit.

How does adiabatic invariance affect the orbital frequency?  Evaluating
\begin{equation}
\delta\Omega = \frac{\partial\Omega}{\partial M}\delta M + 
\frac{\partial\Omega}{\partial S}\delta S + 
\frac{\partial\Omega}{\partial r}\delta r\;,
\end{equation}
with $\delta r$ given by Eq.\ (\ref{eq:circeq_deltar}), we find
\begin{widetext}
\begin{eqnarray}
\frac{\delta\Omega}{\Omega} &=&
\frac{2 - 3v^2 \pm 2av^3(5-9v^2) - 2a^2v^4(3 - 14v^2 + 3v^4) \mp 4a^3v^7(3 - 2v^2) - 3a^4v^{10}}{(1 \pm av^3)^2(1 - 6v^2 \pm 8av^3 - 3a^2v^4)}\frac{\delta M}{M}
\nonumber\\
& & \pm \frac{2v^3(5 - 12v^2) \mp av^4(6 - 33v^2 + 6v^4) - 4a^2v^7(3 - 2v^2) \mp 3a^3v^{10)}}{(1 \pm av^3)^2(1 - 6v^2 \pm 8av^3 - 3a^2v^4)}\frac{\delta S}{M^2}\;.
\label{eq:circeq_deltaOmega}
\end{eqnarray}

\end{widetext}

It is worth contrasting Eq.\ (\ref{eq:circeq_deltaOmega}) with the result that we find if we simply assume that the orbit's radius is fixed during the variation:
\begin{eqnarray}
\frac{\delta\Omega_{\rm wrong}}{\Omega} &=& \frac{1}{\Omega}\left(\frac{\partial\Omega}{\partial M}\delta M + \frac{\partial\Omega}{\partial S}\delta S\right)
\nonumber\\
&=& \frac{(1 \pm 2 av^3)}{2(1\pm av^3)}\frac{\delta M}{M} \mp \frac{v^3}{1 \pm av^3}\frac{\delta S}{M^2}\;.
\label{eq:circeq_deltaOmegawrong}
\end{eqnarray}
As we will discuss in Sec.\ {\ref{sec:inspiral}}, the integrated effect of using Eq.\ (\ref{eq:circeq_deltaOmega}) can differ from the integrated effect of Eq.\ (\ref{eq:circeq_deltaOmegawrong}) by more than an order of magnitude.

Equation (\ref{eq:circeq_deltaOmega}) is a tool that we can use to study how black hole orbits are affected by some process that changes the hole's mass and spin.  Consider, for example, a process which changes the black hole's mass but leaves its spin unchanged --- for example, the infall of mass from far away with zero angular momentum.  The orbital frequencies will change purely by the $\delta M$ term of Eq.\ (\ref{eq:circeq_deltaOmega}).  The $a \to 0$ limit of this is particularly clean:
\begin{equation}
\frac{\delta\Omega}{\Omega} \to \frac{2 - 3v^2}{1 - 6v^2}\frac{\delta M}{M}\;.
\end{equation}
This is consistent with the Newtonian prediction (\ref{eq:NewtonianOmegaChange}), but includes relativistic corrections, including a singularity as we approach the innermost stable orbit.

As another example, consider a process that changes the black hole spin, but does so in a way that leaves the horizon's area unchanged.  Such a change would have $\delta M = \Omega_{\rm H}\delta S$, where $\Omega_{\rm H} = \pm a/2r_{\rm H}$ is the rotation frequency of the hole's event horizon\footnote{$r_{\rm H} = M + M\sqrt{1 - a^2}$ is the coordinate radius of the horizon.}.  By the first law of black hole mechanics, such a process would generate no entropy.  We know of no realistic process which could effect such a change to the black hole's spin, but it is an interesting thought experiment to consider.  The general result we find using this form of $(\delta M,\delta S)$ in Eq.\ (\ref{eq:circeq_deltaOmega}) is fairly lengthy, but the first several terms in the series expansion with $v$ take a simple form:
\begin{equation}
\frac{\delta\Omega}{\Omega} = 10v^3\frac{\delta S}{M^2} + \left(2 + 9v^2 \pm 10av^3\right)\frac{\Omega_{\rm H}\delta S}{M}\;.
\label{eq:circeq_deltaOmega_isentropic}
\end{equation}
The leading terms in $v$ enter at order orbit speed cubed.  As such they, can be significantly less important than terms connected to the black hole's rotation, which enter at order $\Omega_{\rm H}$ (which is itself of order $S$).

Finally, let us imagine a process that changes the mass of the inspiraling body, but leaves the black hole unaffected.  Enforcing
\begin{equation}
\frac{\partial J_\phi}{\partial r}\delta r + \frac{\partial J_\phi}{\partial\mu}\delta\mu = 0
\end{equation}
yields
\begin{equation}
\frac{\delta r}{r} = -\frac{2\left(1 - 3v^2 \pm 2av^3\right)\left(1 \mp 2av^3 + a^2v^4\right)}{\left(1 \pm av^3\right)(\left(1 - 6v^2 \pm 8av^3 - 3a^2v^4\right)}\frac{\delta\mu}{\mu}\;,
\label{eq:circeq_deltar2}
\end{equation}
which in turn yields
\begin{equation}
\frac{\delta\Omega}{\Omega} = \frac{3\left(1 - 3v^2 \pm 2av^3\right)\left(1 \mp 2av^3 + a^2v^4\right)}{\left(1 \pm av^3\right)(\left(1 - 6v^2 \pm 8av^3 - 3a^2v^4\right)}\frac{\delta\mu}{\mu}\;.
\label{eq:circeq_deltaOmega2}
\end{equation}
When $a \to 0$, this yields a result very similar to the Newtonian limit (\ref{eq:NewtonianOmegaChange2}), but with a singularity associated with the approach to the last stable orbit.  Notice that in the case
\begin{equation}
\frac{\delta\Omega}{\Omega} = -\frac{3}{2}\frac{\delta r}{r}\;.
\end{equation}
This follows from the simple dependence of $J_\phi$ on $\mu$, and on the dependence of $\Omega$ on $r$ for circular equatorial orbits.

Equations (\ref{eq:circeq_deltar}), (\ref{eq:circeq_deltaOmega}), (\ref{eq:circeq_deltar2}), and (\ref{eq:circeq_deltaOmega2}) give a complete description for how the orbit of a body about a black hole responds to changes in the mass and spin of the black hole, or to changes to the mass of the orbiting body.  Using these results, it is straightforward to account for how these changes affect quantities which can be observed from such orbits during mass- and spin-changing processes.  In Sec.\ {\ref{sec:inspiral}}, we will examine two examples of such processes in some detail.  Before doing so, we first examine how to account for changes to the properties of more generic black hole orbits.

\section{Black hole orbits II: Weak-field equatorial, slightly eccentric}
\label{sec:eq_wf}

In the following section, we will lay out the framework for applying adiabatic invariance to fully generic Kerr black hole orbits.  To provide some intuition for these results, let us first consider equatorial orbits, looking at the weak-field ($p \gg M$), slightly eccentric ($e \ll 1$) limits.  Tagoshi {\cite{tagoshi}} and Kennefick {\cite{kennefick}} provide the energy and axial angular momentum for such orbits; translating\footnote{The radial parameter $r_0$ used in Refs.\ {\cite{tagoshi,kennefick}} is $p/(1 - e^2)$, the semi-major axis in the elliptic orbit limit.} to the notation we use here, they find
\begin{equation}
E = E_0 + e^2 E_2\;,\quad
L_z = L_{z,0} + e^2 L_{z,2}\;,
\end{equation}
where
\begin{eqnarray}
E_0 &=& \mu\left(\frac{1 - 2u^2 \pm a u^3}{\sqrt{1 - 3u^2 \pm 2au^3}}\right)\;,
\label{eq:kerrenergy0}
\\
L_{z,0} &=& \mu M\left(\frac{1 \mp 2a u^3 + a^2 u^4}{u\sqrt{1 - 3u^2 \pm 2au^3}}\right)\;,
\label{eq:kerrangmom0}
\end{eqnarray}
\begin{widetext}
\begin{eqnarray}
E_2 &=& \mu\frac{u^2\left(2 - 5u^2 \pm au^3 + 2a^2u^4\right)\left(1 - 6u^2 \pm 8au^3 - 3 a^4u^4\right)}{2\left(1 - 2u^2 + a^2u^4\right)\left(1 - 3u^2 \pm 2a u^3\right)^{3/2}}
\label{eq:kerrenergy2}
\\
L_{z,2} &=& \mu M\frac{\left(1 - 6u^2 \pm 8au^3 - 3 a^4u^4\right)\left[1 - 2u^2 \pm au^3\left(1 - 5u^2\right) + a^2u^4\left(2 + u^2\right) \pm 2a^3u^7\right]}{2u\left(1 - 2u^2 + a^2u^4\right)\left(1 - 3u^2 \pm 2a u^3\right)^{3/2}}\;.
\label{eq:kerrangmom2}
\end{eqnarray}
\end{widetext}
In all of these equations, we have put $u = \sqrt{M/p}$, which reduces to $v = \sqrt{M/r}$ (the variable used in the previous section) when $e \to 0$.  Kennefick {\cite{kennefick}} also provides the $O(e^3)$ corrections to $E$ and $L_z$; the $O(e^2)$ terms are sufficient for our purposes.

Using these results, it is straightforward to compute $J_r$ and $J_\phi$ to $O(e^2)$.  The result is
\begin{eqnarray}
J_r &=& \mu M\frac{e^2}{16u}\left[8 + 4u^2 \pm 24au^3 - (37 + 20 a^2)u^4\right]\;,
\\
J_\phi &=& \mu M\frac{1}{16u}\left[16 + 8e^2 + (24 - 12e^2)u^2 \mp 48a(1 - e^2)e^3
\right.
\nonumber\\
& &\left. + \left(54 - 81e^2 + 16a^2(1 - e^2)\right)u^4\right]\;.
\end{eqnarray}
Both of these results are taken to $O(u^3)$, or $O[(M/p)^{3/2}]$, far enough to capture the trends we wish to explore here.

Enforcing adiabatic invariance of $J_r$ and $J_\phi$ as the parameters $\mu$, $M$, and $S$ vary yields
\begin{eqnarray}
\frac{\delta p}{p} &=& -\left[2\left(1 + 3u^2\right) - 18au^3 - e^2u^2\left(4 - 9au\right)\right]\frac{\delta\mu}{\mu}
\nonumber\\
& & - \left[\left(1 + 6u^2\right) - 6au^3 - e^2u^2\left(4 - 3au\right)\right]\frac{\delta M}{M}
\nonumber\\
& &+ 3u^3\left(2 - e^2\right)\frac{\delta S}{M^2}\;,
\\
\frac{\delta e}{e} &=& u^2\left(1 - 9au\right)\frac{\delta\mu}{\mu} + u^2\left(1 - 3au\right)\frac{\delta M}{M} - 3eu^3\frac{\delta S}{M^2}\;.
\nonumber\\
\end{eqnarray}
The $e \to 0$ results for $\delta p/p$ agree with Eq.\ (\ref{eq:circeq_deltar}), expanded to order $v^3$ (using $v = u$ when $e = 0$).  Recall we found $e$ is unchanged for Newtonian orbits when $M$ and $\mu$ are varied.  Given this, it is not surprising that $e$ changes at $O(u^2)$: with factors of $G$ and $c$ restored, $u$ is roughly the orbital speed divided by $c$.  The change we find is consistent with $\delta e$ being a relativistic effect, which suggests that this will be most important for very strong-field black hole orbits.

\section{Generic orbits of an evolving black hole}
\label{sec:kerr_generic}

Turn now to bound generic orbits, the fully relativistic generalization of the Newtonian orbits described in Sec.\ {\ref{sec:newton}}.  Three parameters describe the geometry of such orbits; we use the same set $(p,e,I)$ that we used in Sec.\ {\ref{sec:newton}}.  Outstanding discussion of such orbits can be found in Ref.\ {\cite{schmidt}}.  In particular, Ref.\ {\cite{schmidt}} describes how to compute the constants of the motion $E$, $L_z$, and $Q$ given $(p,e,I)$.  Once those quantities are in hand, it is straightforward to compute the orbit's actions $J_{r,\theta,\phi}$.  Formulas for computing the actions are given in Appendix {\ref{app:actions}}.

As in Sec.\ \ref{sec:kerr_circeq}, we imagine a process that changes the mass of the orbiting body or the mass and spin of the larger black hole, taking $\mu \to \mu + \delta\mu$, $M \to M + \delta M$, and $\quad S \to S + \delta S$.  The orbit's geometry evolves in response, taking
\begin{equation}
p \to p + \delta p\;,\quad e \to e + \delta e\;,\quad I \to I + \delta I
\end{equation}
in such a way that the actions remain fixed:
\begin{equation}
\frac{\partial J_i}{\partial\mu}\,\delta\mu +
\frac{\partial J_i}{\partial M}\,\delta M +
\frac{\partial J_i}{\partial S}\,\delta S +
\frac{\partial J_i}{\partial p}\,\delta p +
\frac{\partial J_i}{\partial e}\,\delta e +
\frac{\partial J_i}{\partial I}\,\delta I = 0
\label{eq:actioninvariant1}
\end{equation}
for $i \in [r,\theta,\phi]$.

\begin{figure}
\includegraphics[width=0.48\textwidth]{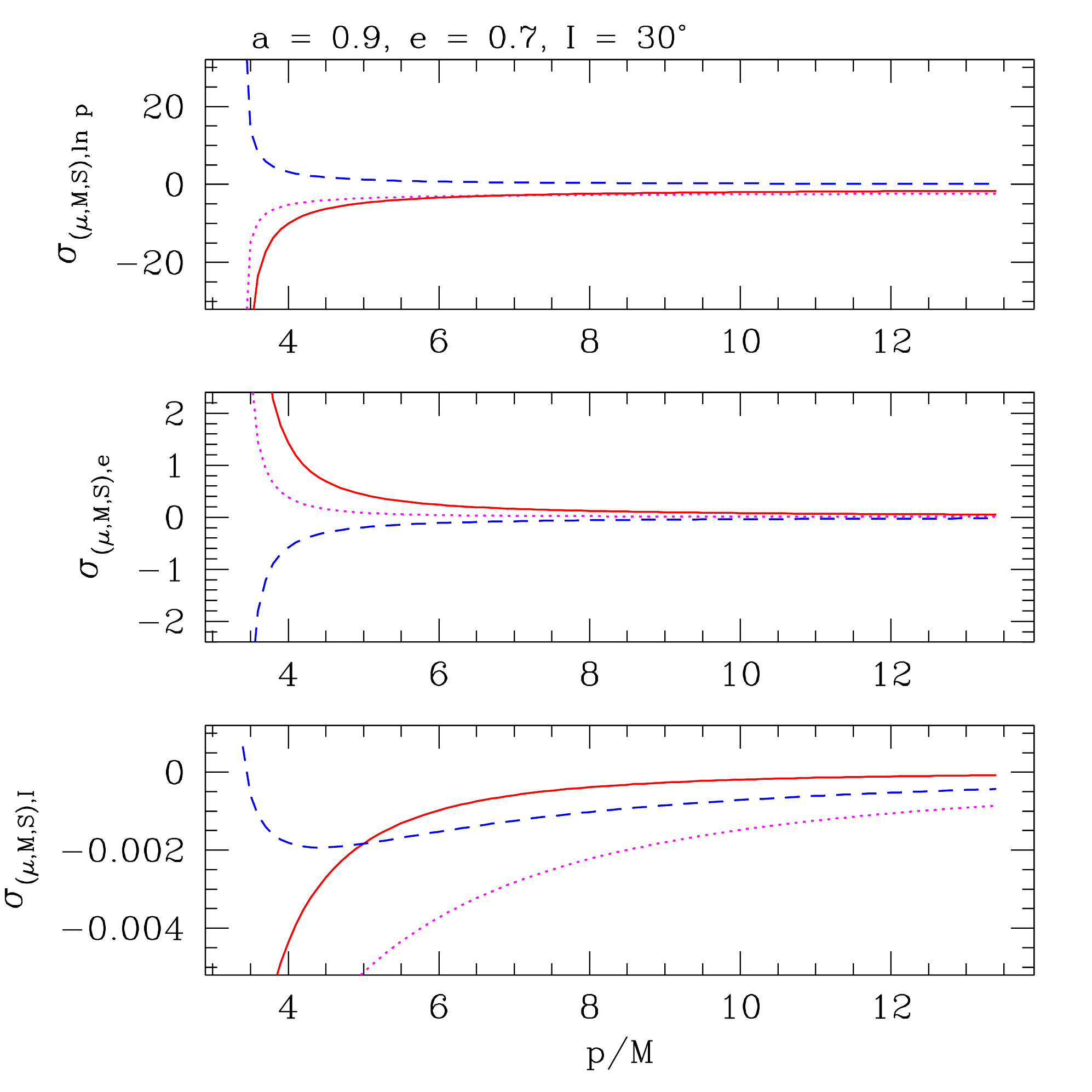}
\caption{Trends for changes in orbit parameters $p$, $e$, and $I$ for a sequence of orbits.  All orbits in the sequence have $a = 0.9$, $e = 0.7$, $I = 30^\circ$; $p$ ranges from just outside the last stable orbit (LSO) to $10M$ beyond the LSO.  The coefficient $\sigma_{y,x}$ describes how orbit parameter $x$ changes per unit increment to binary property $y$; see Eq.\ (\ref{eq:sigmadefs}) for precise definitions of these coefficients.  Top panel shows the change in $\ln p$ (where $p$ is the orbit's semi-latus rectum); middle shows the change in $e$; bottom shows the change in $I$ (in radians).  The change per increment of black hole mass $\delta M$ is plotted in solid red, per increment of black hole spin $\delta S$ in dashed blue, and per increment of orbiting body mass $\delta\mu$ in dotted magenta.  As in the prograde circular equatorial limit [Eq.\ (\ref{eq:circeq_deltar}), upper sign choice], the change per unit $\delta M$ is similar to the change per unit $\delta\mu$; the change per unit $\delta S$ typically is opposite in sign to the mass terms.  This sign difference for the mass and spin terms is expected for orbits with $I < 90^\circ$.  In all cases, the change to the inclination angle is minute.  The examples shown here exhibit the largest changes in $I$ of all the cases that we consider.}
\label{fig:highspin_prograde}
\end{figure}

\begin{figure}
\includegraphics[width=0.48\textwidth]{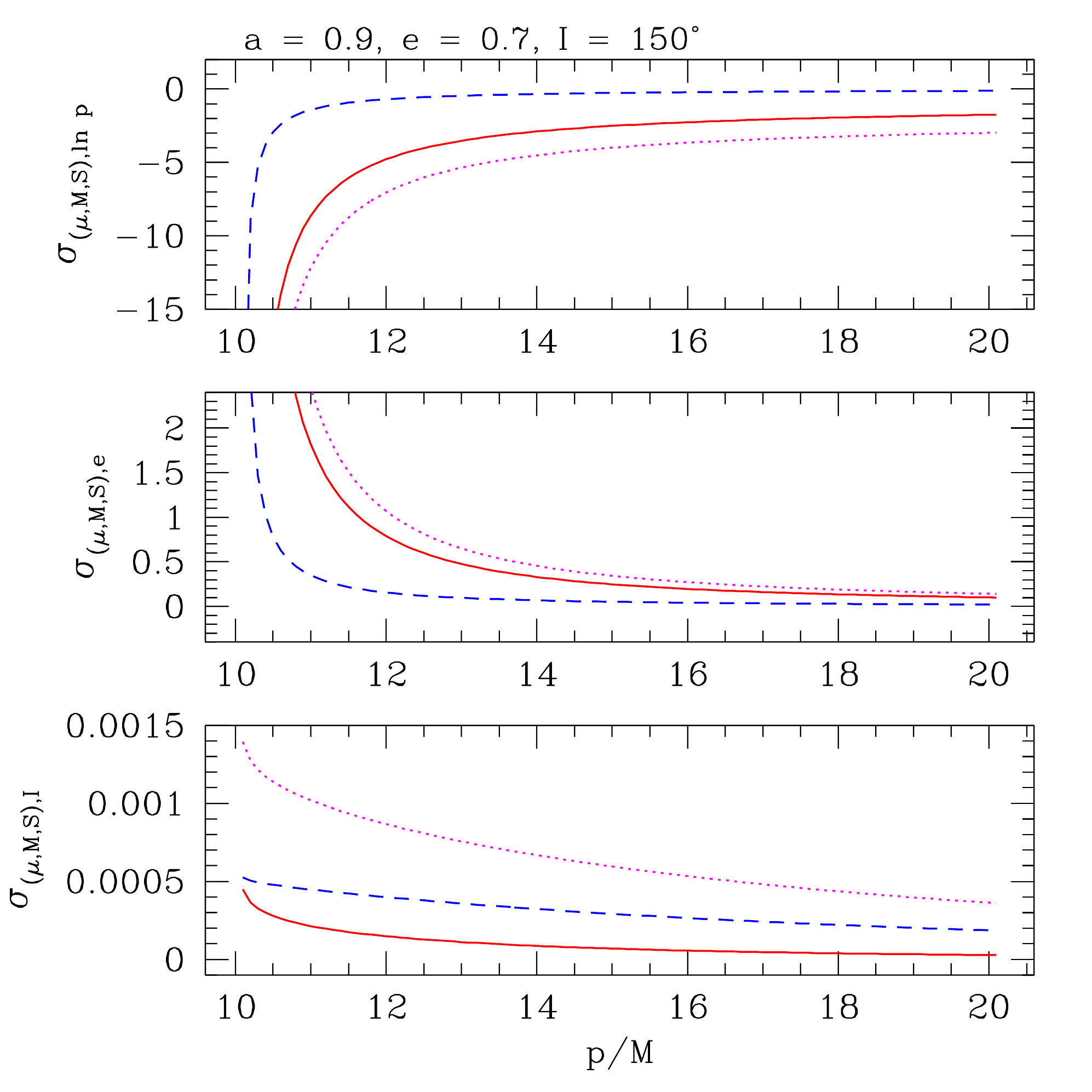}
\caption{Same as Fig.\ {\ref{fig:highspin_prograde}}, but now for a sequence of orbits whose inclination angle is $I = 150^\circ$.  Many of the trends noted in the caption to Fig.\ {\ref{fig:highspin_prograde}} are reflected here as well.  In this case, the mass and spin terms have the same sign, which is consistent with what we found in the retrograde circular equatorial case [Eq.\ (\ref{eq:circeq_deltar}), lower sign choice].  We expect this behavior for orbits with $I > 90^\circ$.}
\label{fig:highspin_retrograde}
\end{figure}

\begin{figure}
\includegraphics[width=0.48\textwidth]{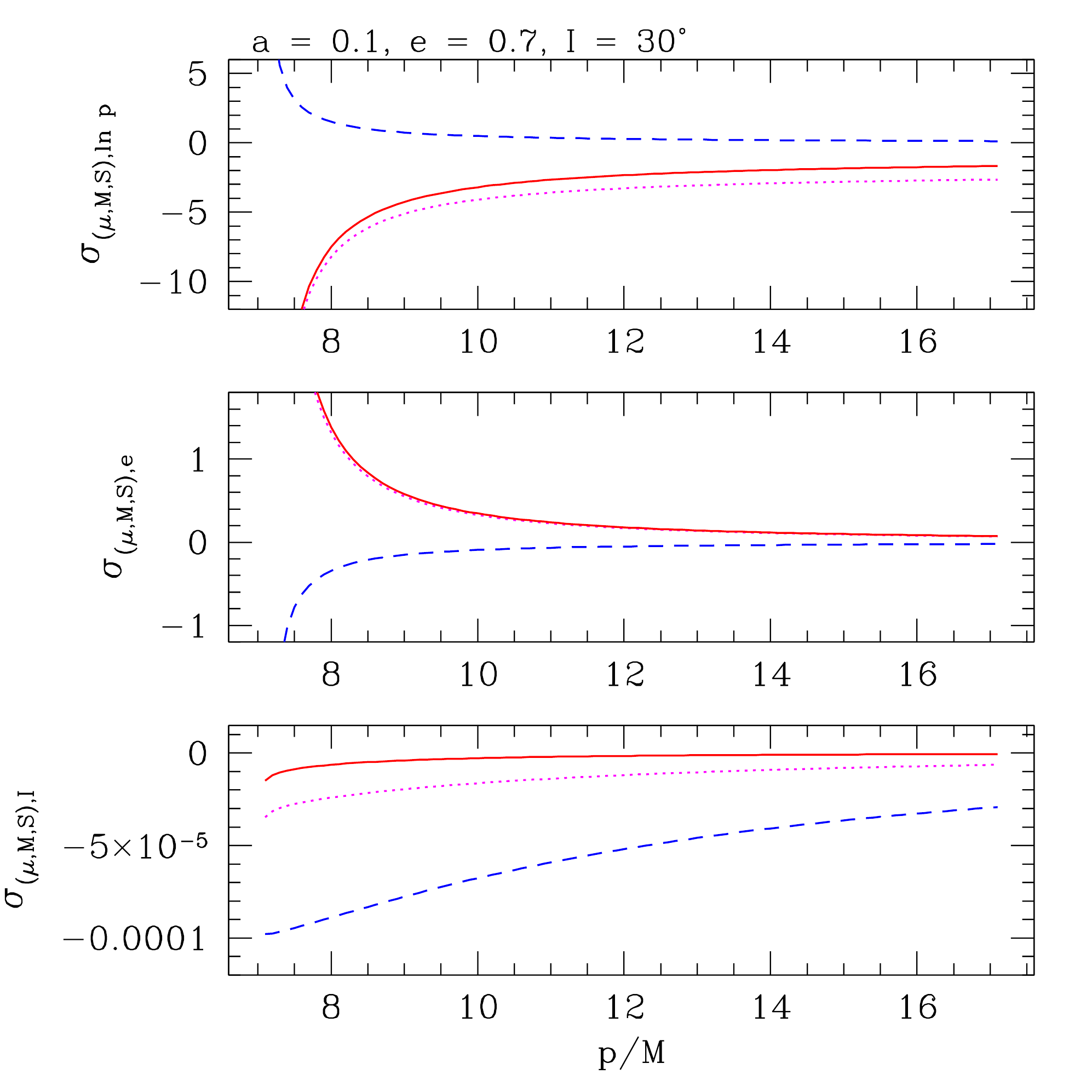}
\caption{Same as Fig.\ {\ref{fig:highspin_prograde}}, but now for a sequence of orbits about a black hole with $a = 0.1$.  The behavior of changes to $p$ and $e$ is similar to that seen for the high-spin sequence.  Perhaps most noteworthy here is that the change to the orbit's inclination is even smaller in this case, consistent with the nearly spherical nature of the spacetime for such a slowly spinning black hole.}
\label{fig:lowspin_prograde}
\end{figure}

Let us write Eq.\ (\ref{eq:actioninvariant1}) as a matrix equation: we put
\begin{equation}
\mathsf{J}\cdot\mathsf{\delta O} = -\mathsf{\delta H}\;,
\end{equation}
where $\mathsf{J}$ is the matrix of action derivatives,
\begin{equation}
\mathsf{J} = 
\begin{pmatrix}
\partial J_r/\partial p &
\partial J_r/\partial e &
\partial J_r/\partial I \cr
\partial J_\theta/\partial p &
\partial J_\theta/\partial e &
\partial J_\theta/\partial I \cr
\partial J_\phi/\partial p &
\partial J_\phi/\partial e &
\partial J_\phi/\partial I
\end{pmatrix}\;,
\end{equation}
the vector $\delta\mathsf{O}$ represents changes to the orbit's geometry,
\begin{equation}
\delta\mathsf{O} = 
\begin{pmatrix}
\delta p \cr \delta e \cr \delta I
\end{pmatrix}\;,
\end{equation}
and the vector $\delta\mathsf{H}$ represents changes in the actions due to variations in the black hole's or the orbiting body's properties,
\begin{equation}
\delta\mathsf{H} = 
\begin{pmatrix}
(\partial J_r/\partial\mu)\delta\mu +
(\partial J_r/\partial M)\delta M +
(\partial J_r/\partial S)\delta S\cr
(\partial J_\theta/\partial\mu)\delta\mu +
(\partial J_\theta/\partial M)\delta M +
(\partial J_\theta/\partial S)\delta S\cr
(\partial J_\phi/\partial\mu)\delta\mu +
(\partial J_\phi/\partial M)\delta M +
(\partial J_\phi/\partial S)\delta S
\end{pmatrix}\;.
\end{equation}
The action derivatives can all be computed by simple quadratures; see Appendix {\ref{app:actions}} for discussion and relevant formulas.  With $\mathsf{J}$ and $\delta\mathsf{H}$ computed, we then have
\begin{equation}
\delta\mathsf{O} = -\mathsf{J}^{-1}\cdot\delta\mathsf{H}\;.
\label{eq:generic_orbitevolve}
\end{equation}
Given orbit parameters, it is straightforward to numerically solve Eq.\ (\ref{eq:generic_orbitevolve}).  The solutions we find take the form
\begin{equation}
\delta x = \sigma_{\mu,x} \frac{\delta\mu}{\mu} + \sigma_{M,x} \frac{\delta M}{M} + \sigma_{S,x} \frac{\delta S}{M^2}\;,
\label{eq:sigmadefs}\\
\end{equation}
for $x \in (\ln p, e, I)$.  (We use $\ln p$ since the solutions we find are best expressed using $\delta\ln p = \delta p/p$.)  Representative examples of $\sigma_{(\mu,M,S),x}$ are shown in Figs.\ \ref{fig:highspin_prograde} -- \ref{fig:lowspin_prograde}.

A summary of the general trends we find is:

\begin{itemize}

\item For the change in $p$, we find the coefficient of the mass terms, $\sigma_{M,{\ln p}}$ and $\sigma_{\mu,{\ln p}}$, to be of order unity away from the last stable orbit (LSO), but to diverge as the LSO is approached.  The spin term $\sigma_{S,{\ln p}}$ is of order $0.1 - 0.2$ at large $p$, also diverging near the LSO.  The spin term is positive if the orbit is prograde ($L_z > 0$, $I \le 90^\circ$), and is negative otherwise.

\item The mass terms $\sigma_{M,e}$ and $\sigma_{\mu,e}$ tend to make orbits more eccentric.  The corresponding spin term $\sigma_{S,e}$ decreases eccentricity for prograde orbits, and increases it for retrograde, with both terms also diverging at the LSO.

\item The change in inclination is always minute.  For orbits of Schwarzschild black holes, $\delta I$ vanishes entirely (in keeping with the spherical nature of the spacetime).  Even for rapid spin, the magnitude of the change tends to be rather paltry.

\end{itemize}

\section{Impact on inspiral}
\label{sec:inspiral}

We have shown that if a black hole slowly evolves, changing from a Kerr solution with mass and spin $(M, S)$ to another Kerr solution with $(M + \delta M, S + \delta S)$, or the orbiting body's mass changes from $\mu$ to $\mu + \delta\mu$, then adiabatic invariance demands that orbits of that black hole will change in order for the orbits' actions to remain constant.  At least in principle, this could have observational consequences, since the change to the orbit affects characteristics like orbital frequencies.  We have seen from some simple examples that neglecting the change to the orbit that is due to adiabatic invariance can underestimate how orbital frequencies change due to the hole's mass and spin evolution.

Is this effect important in practice?  To test this, let us apply this idea to two situations that have been analyzed in past work.  In both cases, we will examine an extreme mass-ratio inspiral, focusing on the relatively simple circular and equatorial limit.  In the first case, we will examine how the orbital phase of the inspiral is changed if the mass and spin of the black hole changes due to accretion.  We will ignore the effect of viscous drag due to this accreting material --- an unrealistic approximation, but one that allows us to isolate and quantify how the effect we study in this paper impacts binary evolution.  In the second case, we we will examine how the inspiral is changed if we account for energy and angular momentum carried by radiation into the black hole.

Both of these effects have been studied in past work (for example, {\cite{bcp14}} includes this aspect of accretion physics in their very comprehensive study of environmental effects on gravitational-wave emitting binaries, and {\cite{in2018}} includes the change to the black hole properties that comes from absorbed radiation).  The impact of these effects has been found to be quite small or even negligible.  However, in both cases it was assumed that the orbit's geometry was fixed as the black hole evolved, an assumption that is inconsistent with the adiabatic invariance of the orbit's actions.  As we will now show, it remains the case that the integrated impact of these effects is small, but not quite as small as past work found.

It is worth emphasizing that the impact of adiabatic invariance must be separated from the impact of other effects on the system.  For example, the backreaction of gravitational waves does not evolve the system in a way that leaves the actions unchanged.  The extreme mass-ratio limit is an excellent laboratory for studying the importance of the actions' adiabatic invariance because it allows this separation to be made very cleanly.  In essence, we assume that for every short time interval $\delta t$ over which the system evolves, we separately allow the system to evolve due to gravitational-wave backreaction, and then due to changes of black hole's mass and spin.  As long as the impact of all these effects are small over each interval, and all aspects of the system change adiabatically (i.e., that $|\dot X\,\delta t| \ll |X|$ for any $X$ that characterizes the system), this is a reasonable assumption to make.

Let us begin by first writing down in general terms how we will quantify the impact of the black hole's evolution.  Our main work horse is Eq.\ (\ref{eq:circeq_deltaOmega}), which we write
\begin{equation}
\delta\Omega = \Omega\left(\sigma_{M,\Omega}\frac{\delta M}{M} + \sigma_{S,\Omega}\frac{\delta S}{M^2}\right)\;.
\label{eq:deltaOmega_forestimate}
\end{equation}
We will also examine how this binary evolves using Eq.\ (\ref{eq:circeq_deltaOmegawrong}), the incorrect change in $\Omega$ found by neglecting the change in the orbit that comes from enforcing adiabatic invariance.

We will assume that some process changes the black hole's mass and spin at rates $(dM/dt,dS/dt)$.  Combining this with Eq.\ (\ref{eq:deltaOmega_forestimate}) tells us that the frequency shifts away from $\Omega$ at the rate
\begin{equation}
\frac{d\delta\Omega}{dt} = \Omega\left(\frac{\sigma_{M,\Omega}}{M}\frac{dM}{dt} + \frac{\sigma_{S,\Omega}}{M^2}\frac{dS}{dt}\right)\;.
\end{equation}
The total accumulated frequency shift measured over an interval $t_{\rm S} \le t \le t_{\rm F}$ is given by integrating this up:
\begin{equation}
\Delta\Omega = \int_{t_{\rm S}}^{t_{\rm F}}\left(\frac{\sigma_{M,\Omega}}{M}\frac{dM}{dt} + \frac{\sigma_{S,\Omega}}{M^2}\frac{dS}{dt}\right)\Omega\,dt\;.
\end{equation}
To compute the orbital phase associated with these effects, we integrate again:
\begin{eqnarray}
\Delta\Phi &=& \int_{t_{\rm S}}^{t_{\rm F}}\Delta\Omega(t)\,dt
\nonumber\\
&=& \int_{t_{\rm S}}^{t_{\rm F}}\left[\int_{t_{\rm S}}^t\left(\frac{\sigma_{M,\omega}}{M}\frac{dM}{dt} + \frac{\sigma_{S,\Omega}}{M^2}\frac{dS}{dt}\right)\Omega\,dt'\right]dt\;.
\nonumber\\
\label{eq:DeltaPhi1}
\end{eqnarray}

All of the quantities which appear under the integrals in Eq.\ (\ref{eq:DeltaPhi1}) are more naturally expressed as functions of orbital radius than as functions of time, so it is more convenient to cast this as an integral over orbital radius.  Assuming that the inspiral is driven by gravitational-wave emission, we have
\begin{equation}
dt = \left(\frac{dE^{\rm orb}/dr}{\dot E^\infty + \dot E^{\rm H}}\right)dr\;.
\end{equation}
Here, $E^{\rm orb}$ is the orbital energy as a function of $r$, which is given by Eq.\ (\ref{eq:kerrenergy0}) with $u \to v \equiv \sqrt{M/r}$.  The quantities $\dot E^{\infty,{\rm H}}$ are the fluxes of energy carried by gravitational waves to infinity and down the horizon; a high-order fit to these quantities is provided by Ref.\ {\cite{fujita2015}}, with the terms that we use given in Appendix {\ref{app:flux}}.  The change to the orbital phase accumulated over an inspiral from $r_{\rm S}$ to the innermost stable circular orbit, $r_{\rm ISCO}$, is given by
\begin{widetext}
\begin{equation}
\Delta\Phi = \int_{r_{\rm S}}^{r_{\rm ISCO}}\left[\int_{r_{\rm S}}^r
\left(\frac{\sigma_{M,\omega}}{M}\frac{dM}{dt} + \frac{\sigma_{S,\Omega}}{M^2}\frac{dS}{dt}\right)\Omega\left(\frac{dE^{\rm orb}/dr}{\dot E^\infty + \dot E^{\rm H}}\right)dr'\right]\left(\frac{dE^{\rm orb}/dr}{\dot E^\infty + \dot E^{\rm H}}\right)dr\;.
\label{eq:DeltaPhi2}
\end{equation}
\end{widetext}
Formulae for $r_{\rm ISCO}$ can be found in Ref.\ {\cite{bpt72}}.  In Eq.\ (\ref{eq:DeltaPhi2}), all of the quantities in the inner integral (in square brackets) are taken to be functions of $r'$; quantities in the outer integral are taken to be functions of $r$.

We now examine Eq.\ (\ref{eq:DeltaPhi2}) for the two cases that we consider.

\subsection{Mass and spin evolution due to accretion}
\label{eq:deltaphi_accretion}

Let us assume that the black hole's mass is changing at the rate
\begin{equation}
\frac{dM}{dt} = 2\times10^{-3}\left(\frac{M}{10^6\,M_\odot}\right)\,M_\odot\;{\rm yr}^{-1}\;,
\label{eq:MdotEdd}
\end{equation}
which is approximately one tenth of the Eddington rate assuming $10\%$ radiative efficiency.  Let us assume that each infalling mass element carries into the black hole the angular momentum of an orbit at the innermost stable circular orbit, so that
\begin{equation}
\frac{dS}{dt} = \frac{1}{\Omega_{\rm ISCO}}\frac{dM}{dt}\;,
\end{equation}
The orbital frequency at this orbit can be found by evaluating Eq.\ (\ref{eq:circeq_Omega}) at $r_{\rm ISCO}$.  We emphasize that these forms for $dM/dt$ and $dS/dt$ are not intended to model any realistic accretion scenario.  We take the large black hole to have mass $M = 10^6\,M_\odot$, the inspiraling body to have $\mu = 10\,M_\odot$, and begin the inspiral at $r_{\rm S} = 10M$.

\begin{figure}
\includegraphics[width=0.48\textwidth]{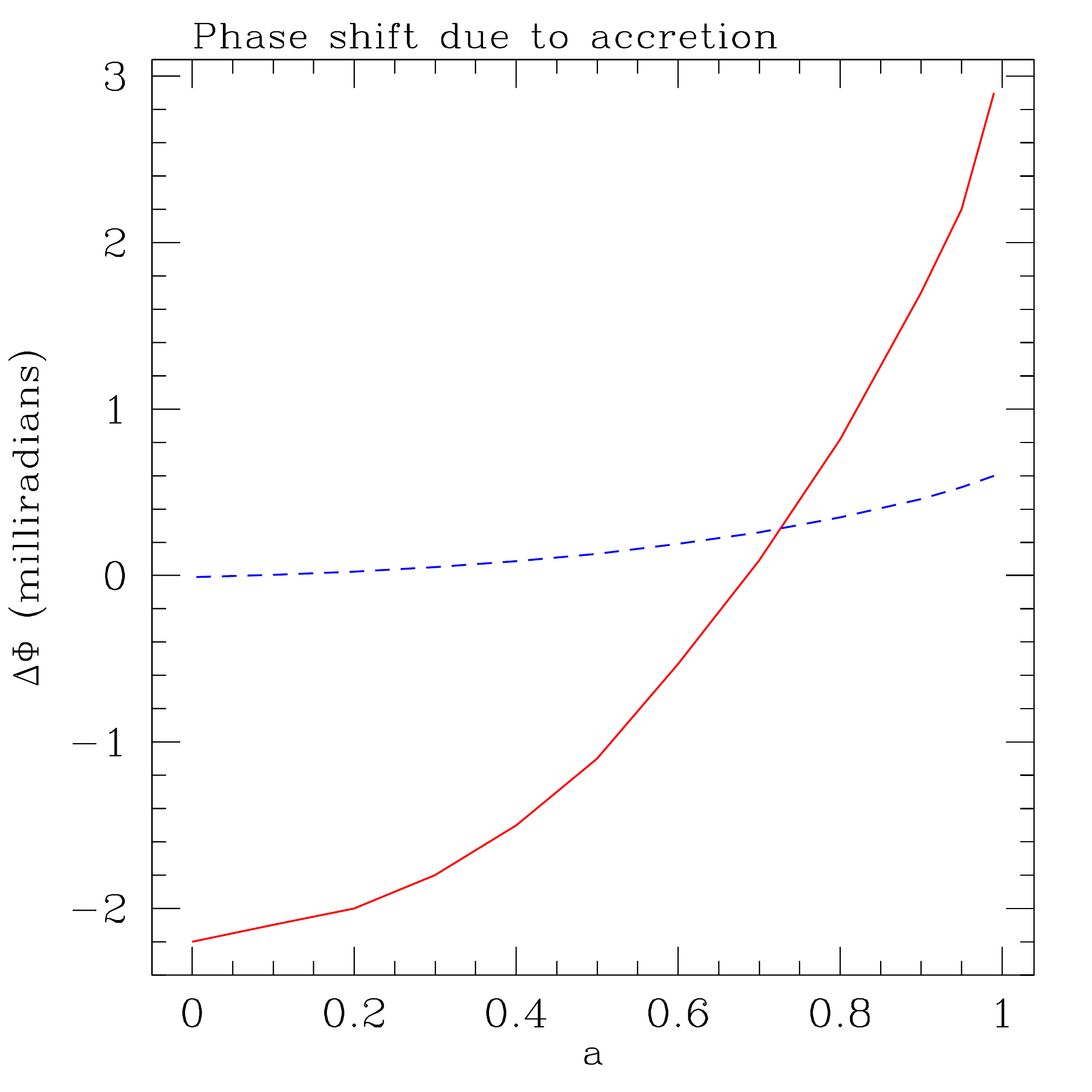}
\caption{The phase shift $\Delta\Phi$ arising from changes to a black hole's mass and spin due to accretion, accumulated during a prograde circular, equatorial inspiral from $r_{\rm S} = 10M$ to the innermost stable circular orbit.  We assume a $10\,M_\odot$ small body spiraling into a $10^6\,M_\odot$ black hole, we assume that mass accretes at roughly $10\%$ of the Eddington rate, and each element carries its angular momentum at the innermost stable circular orbit into the horizon.  The solid (red) curve shows the phase shift that accumulates properly taking into account adiabatic invariance of the actions; the dashed (blue) curve neglects this physics, changing the black hole's mass and spin but assuming the orbit's geometry is unaffected.  Although both phase shifts are quite small, notice the important impact of properly accounting for adiabatic invariance.}
\label{fig:deltaphi_acc}
\end{figure}

Figure {\ref{fig:deltaphi_acc}} shows the result of this calculation for prograde inspiral, plotting $\Delta\Phi$ for a range of starting black hole spins.  We show both the result found by enforcing for adiabatic invariance, as well as the result one would find simply holding the orbit at fixed radius as the hole's mass and spin secularly evolve.  Both shifts are small, but the trends we find are quite different --- properly accounting for adiabatic invariance has a significant impact.  It is worth noting that in this case the phase shift scales with the system masses as $(M/\mu)^2$; the small body spends more time spiraling through the integration interval at more extreme mass ratio.  One must do this analysis with a more realistic accretion model than the {\it ad hoc} scenario that we have constructed, but it is interesting that the phase shift would be not quite so paltry for more extreme mass ratios.  

\subsection{Mass and spin evolution due to absorbed radiation}

Next let us consider the impact of accounting for how the black hole's mass and spin due to radiation absorbed by the black hole.  We use
\begin{equation}
\frac{dM}{dt} = -\dot E^{\rm H}\;,\quad
\frac{dS}{dt} = -\frac{1}{\Omega}\dot E^{\rm H}\;.
\end{equation}
The minus signs in these expressions enforce global conservation: the mass of the black hole increases due to the loss of energy from the orbit, and likewise for the spin of the black hole.

\begin{figure}
\includegraphics[width=0.48\textwidth]{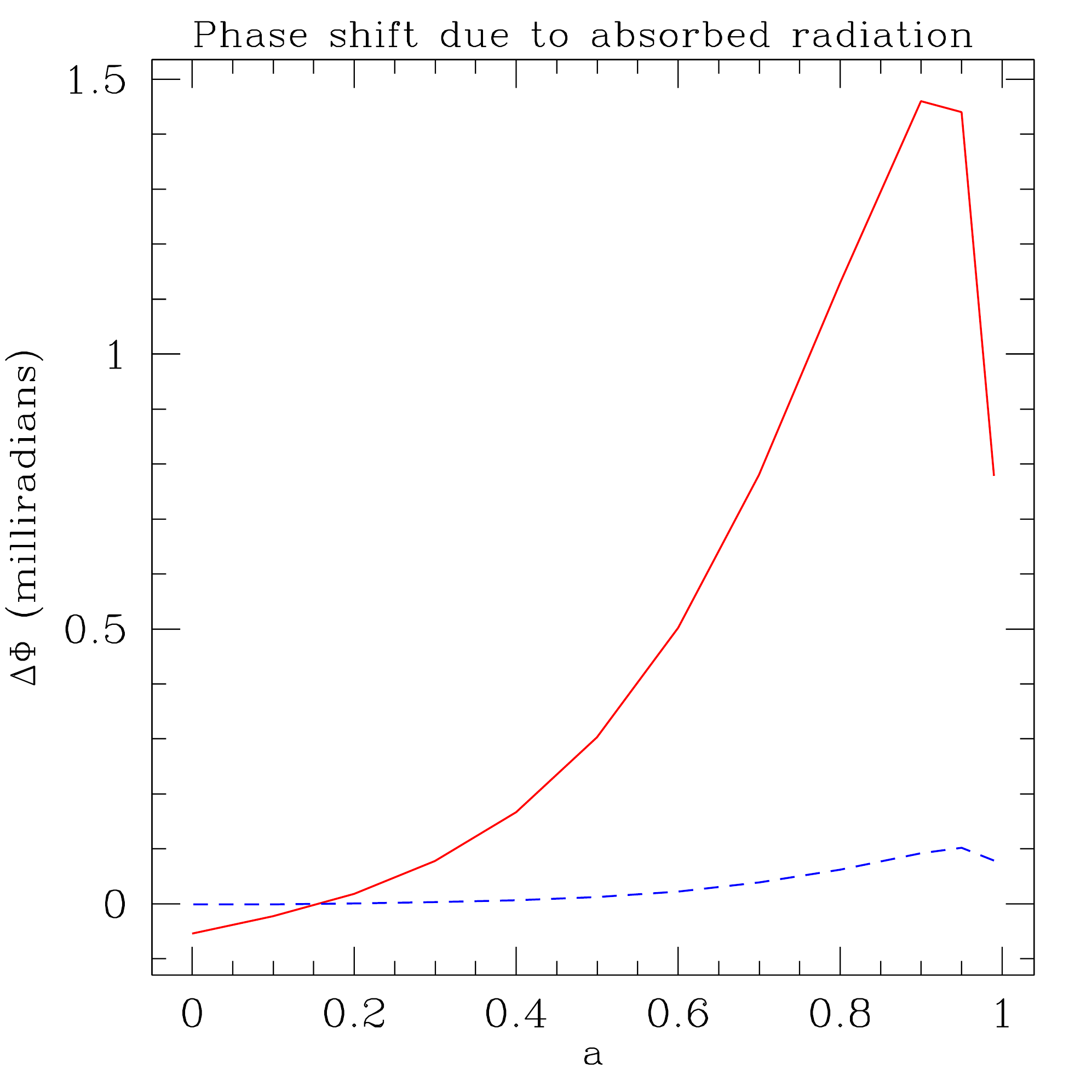}
\caption{The phase shift $\Delta\Phi$ arising from changes to a black hole's mass and spin due to radiation absorbed from the orbit, accumulated during a prograde circular, equatorial inspiral from $r_{\rm S} = 10M$ to the innermost stable circular orbit.  This result is independent of mass ratio (provided the mass ratio is extreme enough that the large mass-ratio formulas we use are accurate).  The solid (red) curve shows the phase shift that accumulates with adiabatic invariance properly taken into account; the dashed (blue) curve shows the result found when these effects are neglected.  Although both phase shifts are extremely small, the curve which accounts for adiabatic invariance is larger by a factor of 10 -- 20 across a wide range of spin.}
\label{fig:deltaphi_rad}
\end{figure}

Figure {\ref{fig:deltaphi_rad}} shows the resulting phase shift we find.  We again show both the result enforcing adiabatic invariance, as well as the result one would find if the orbit were held at fixed orbital radius.  As expected, the phase shift is quite small: across all the spins that we have examined, it is of order milliradians at most.  This is essentially consistent with the discussion in Ref.\ {\cite{in2018}}, which finds that including the secular evolution of black hole mass and spin has a puny effect on measurement templates.  However, it must be noted that not accounting for adiabatic invariance can lead to an underestimate of the shift by a factor of 10 -- 20.  The correct phase shift is not quite as puny as past work has found.

\section{Conclusion}
\label{sec:conclude}

Not surprisingly, the integrated effect of how inspiral into a black hole is modified by evolution of the hole's mass and spin is quite small.  Indeed, the milliradian-level phase shift we find is unlikely to be of observational significance.  A rough rule of thumb is that a template phase accuracy of ``a fraction of a radian divided by the signal to noise ratio'' is needed to insure that systematic errors (due to mismodeling, for example) are smaller than statistical errors (due to noise) (see, e.g., Ref.\ {\cite{lob08}}).  It is conceivable that the effect we find may touch observational significance for some accretion scenarios, depending on system mass ratio and the properties of the accretion flow.  It is highly unlikely that the effect of absorbed radiation will be of observational significance, except conceivably for sources with extremely high signal-to-noise ratio.

It is interesting to note that the second effect impacts the system at second order in the system's mass ratio, and as such should emerge in an appropriately averaged self-force analysis.  The shift to the spacetime $(\delta M, \delta S)$ and the shift to the orbit $(\delta p, \delta e, \delta I)$ can be recast as a shift to the orbit integrals $(\delta E, \delta Q)$.  (Note that $\delta L_z = 0$, since $L_z \equiv J_\phi$ is itself an adiabatic invariant.)  The analysis we have presented here may serve to provide a simple calculation for a quantity which, though small, may serve as a useful check for the community as self force calculations are pushed to higher order and are applied to astrophysically realistic extreme mass-ratio systems.

\acknowledgments

This work was supported by NSF grant PHY-1707549.  We thank Zoltan Haiman, who suggested examining variations to the smaller body's mass, and various participants of the 2018 Capra Meeting in Golm, Germany (particularly Adam Pound) regarding order counting and the integrated impact of the absorbed radiation on binary inspiral.  We are especially grateful to Deepto Chakrabarty for not reclaiming the copy of Binney and Tremaine that he loaned to us in early 2017, and promise to return it when a version of this paper survives peer review.

\appendix

\section{A proof of adiabatic invariance}
\label{app:proof}

In this appendix, we give a brief proof that the actions $J_k$ are invariant under slow evolution of the potential or spacetime in which they are defined.  This proof very closely follows that given in Ref.\ {\cite{bt1987}}; see in particular discussion in their Chapter 3.6 and their Appendix I.D4.

Consider an integrable dynamical system in $N$ spatial dimensions.  Assume that this system can be represented using canonical coordinates ${\bf q} = (q^1, q^2, \ldots, q^N)$, with conjugate momenta ${\bf p} = (p_1, p_2, \ldots, p_N)$.  This system presumably represents motion in some potential $V({\bf x})$ or some spacetime $g_{\mu\nu}({\bf r})$.  For the purpose of this appendix, we are agnostic about the precise nature of the system.

Let $\gamma_k(t)$ be a particular 1-dimensional closed trajectory through the $2N$-dimensional phase space $({\bf q}, {\bf p})$ at time $t$.  We can choose $N$ such trajectories in such a way that the systems's $N$ action variables can be written
\begin{equation}
J_k(t) = \frac{1}{2\pi}\oint_{\gamma_k(t)} {\bf p}\cdot d{\bf q}\;.
\label{eq:action1}
\end{equation}
Let ${\cal S}_k(t)$ be a 2-surface in phase space that has $\gamma_k(t)$ as its boundary.  Using Green's theorem, Eq.\ (\ref{eq:action1}) can be rewritten
\begin{equation}
J_k(t) = \frac{1}{2\pi}\int_{{\cal S}_k(t)} d{\bf p}\cdot d{\bf q} = \frac{1}{2\pi}\int_{{\cal S}_k(t)} dp_i dq^i\;.
\label{eq:action2}
\end{equation}
We use the summation convention here: repeated indices in the upstairs and downstairs position are assumed to be summed from $1$ to $N$.  Let us define coordinates $(u,v)$ which cover the surface ${\cal S}_k(t)$.  We can then write the action
\begin{equation}
J_k(t) = \frac{1}{2\pi}\int_{{\cal S}_k(t)} \frac{\partial(p_i, q^i)}{\partial(u,v)}du\,dv\;,
\end{equation}
where we have introduced the Jacobian between $(u,v)$ and the canonical coordinates and momenta:
\begin{equation}
\frac{\partial(p_i,q^i)}{\partial(u,v)} = \frac{\partial p_i}{\partial u}\frac{\partial q^i}{\partial v} - \frac{\partial p_i}{\partial v}\frac{\partial q^j}{\partial u}\;.
\end{equation}

Let us next examine how these quantities evolve in time.  We assume that the system evolves from one integrable configuration to another, which means that the system evolves via a time-dependent Hamiltonian ${\bf H}$ such that after a time interval $\delta t$
\begin{eqnarray}
p_i &\to& p_i = p_i - \frac{\partial{\bf H}}{\partial q^i}\delta t\;,
\label{eq:evolve_p}\\
q^i &\to& {q'}^i = q^i + \frac{\partial{\bf H}}{\partial p_i}\delta t\;,
\label{eq:evolve_q}
\end{eqnarray}
plus corrections of $O(\delta t^2)$.  Each point $(u,v)$ on ${\cal S}_k(t)$ is associated with a point in phase space $[{\bf p}(u,v;t),{\bf q}(u,v;t)]$; under the action of ${\bf H}$, the point $(u,v)$ becomes associated with a new point $[{\bf p}(u,v; t+ \delta t),{\bf q}(u,v; t + \delta t)]$.  Hence, under the action of ${\bf H}$, the surface ${\cal S}_k(t)$ evolves to a new surface ${\cal S}_k(t + \delta t)$ with boundary $\gamma_k(t + \delta t)$, but spanned by the same range of the coordinates $(u,v)$.  In other words, ${\bf H}$ maps each point $(u,v)$ on ${\cal S}_k(t)$ to a corresponding point $(u,v)$ on ${\cal S}_k(t + \delta t)$.

Let us now compute how the actions change under this operation:
\begin{eqnarray}
\frac{dJ_k}{dt} &=& \lim_{\delta t \to 0}\frac{1}{\delta t}\left[J_k(t + \delta t) - J_k(t)\right]
\nonumber\\
&=& \lim_{\delta t\to0}\frac{1}{\delta t}\frac{1}{2\pi}\Biggl[\int_{{\cal S}_k(t + \delta t)} \frac{\partial(p_i',{q'}^i)}{\partial(u,v)} du\,dv
\nonumber\\
& &\qquad\qquad- \int_{{\cal S}_k(t)}\frac{\partial(p_i,q^i)}{\partial(u,v)}du\,dv\Biggr]
\nonumber\\
&=& \lim_{\delta t\to0}\frac{1}{\delta t}\frac{1}{2\pi}\int_{{\cal S}_k} \left[\frac{\partial(p_i',{q'}^i)}{\partial(u,v)}  - \frac{\partial(p_i,q^i)}{\partial(u,v)}\right]du\,dv\;.
\nonumber\\
\label{eq:dJkdt1}
\end{eqnarray}
Two aspects of Eq.\ (\ref{eq:dJkdt1}) are worth comment.  As discussed in the text following Eq.\ (\ref{eq:evolve_q}), the surfaces ${\cal S}_k(t)$ and ${\cal S}_k(t + \delta t)$ are covered by the same range of coordinates $(u,v)$ thanks to how the Hamiltonian ${\bf H}$ maps $({\bf p}, {\bf q}) \to ({\bf p'}, {\bf q'})$.  As such, either ${\cal S}_k(t)$ or ${\cal S}_k(t + \delta t)$ can be used to express the range of integration; we have left off the time dependence in ${\cal S}_k$ on the final line of Eq.\ (\ref{eq:dJkdt1}) to express this.

Second, note that we assume the system's motion is such that the system passes through all points in both the surface ${\cal S}_k(t)$ and ${\cal S}_k(t + \delta t)$.  This is where the requirement that the system is ``slowly evolving'' enters our analysis.  If the system is evolving so rapidly that this assumption is not correct, then we are violating this requirement and the analysis here is not applicable to this problem.  We comment on this point further at this end of this appendix.

Continuing our calculation, let us write out the Jacobian between $(p'_i, {q'}^i)$ and $(u,v)$: using Eqs.\ (\ref{eq:evolve_p}) and (\ref{eq:evolve_q}), we have
\begin{eqnarray}
\frac{\partial(p_i',{q'}^i)}{\partial(u,v)} &=& \left(\frac{\partial p_i}{\partial u} - \frac{\partial^2{\bf H}}{\partial u\partial q^i}\delta t\right)\left(\frac{\partial q^i}{\partial v} - \frac{\partial^2{\bf H}}{\partial v\partial p_i}\delta t\right)\nonumber\\
&-& \left(\frac{\partial p_i}{\partial v} - \frac{\partial^2{\bf H}}{\partial v\partial q^i}\delta t\right)\left(\frac{\partial q^i}{\partial u} - \frac{\partial^2{\bf H}}{\partial u\partial p_i}\delta t\right)\;,
\nonumber\\
\end{eqnarray}
from which we see that
\begin{eqnarray}
& &\frac{\partial(p_i',{q'}^i)}{\partial(u,v)} - \frac{\partial(p_i,q^i)}{\partial(u,v)} =
\nonumber\\
& &\quad\delta t\Biggl(\frac{\partial p_i}{\partial u}\frac{\partial^2{\bf H}}{\partial p_i\partial v} - \frac{\partial q^i}{\partial v}\frac{\partial^2{\bf H}}{\partial q^i\partial u} + \frac{\partial q^i}{\partial u}\frac{\partial^2{\bf H}}{\partial q^i\partial v} - \frac{\partial p_i}{\partial v}\frac{\partial^2{\bf H}}{\partial p_i\partial u}\Biggr)
\nonumber\\
& &\quad + O(\delta t^2)\;.
\label{eq:Jacobiandiff}
\end{eqnarray}
The Hamiltonian is most naturally expressed in terms of $(p_i, q^i)$, so it is useful to expand the derivatives of ${\bf H}$ in Eq.\ (\ref{eq:Jacobiandiff}) using
\begin{eqnarray}
\frac{\partial {\bf H}}{\partial u} &=& \frac{\partial p_j}{\partial u}\frac{\partial {\bf H}}{\partial p_j} + \frac{\partial q^j}{\partial u}\frac{\partial {\bf H}}{\partial q^j}\;,
\nonumber\\
\frac{\partial {\bf H}}{\partial v} &=& \frac{\partial p_j}{\partial v}\frac{\partial {\bf H}}{\partial p_j} + \frac{\partial q^j}{\partial v}\frac{\partial {\bf H}}{\partial q^j}\;.
\label{eq:Hamiltonianderiv}
\end{eqnarray}
Combining Eqs.\ (\ref{eq:dJkdt1}), (\ref{eq:Jacobiandiff}), and (\ref{eq:Hamiltonianderiv}) we find
\begin{eqnarray}
\frac{dJ_k}{dt} &=& \frac{1}{2\pi}\int_{{\cal S}_k} du\,dv\Biggl(\frac{\partial p_i}{\partial u}\frac{\partial q^j}{\partial v}\frac{\partial^2{\bf H}}{\partial p_i\partial q^j} +
\frac{\partial p_i}{\partial u}\frac{\partial p_j}{\partial v}\frac{\partial^2{\bf H}}{\partial p_i\partial p_j}
\nonumber\\
& & - \frac{\partial q^i}{\partial v}\frac{\partial q^j}{\partial u}\frac{\partial^2{\bf H}}{\partial q^i\partial q^j} - \frac{\partial q^i}{\partial v}\frac{\partial p_j}{\partial u}\frac{\partial^2{\bf H}}{\partial q^i\partial p_j}
\nonumber\\
& & + \frac{\partial q^i}{\partial u}\frac{\partial q^j}{\partial v}\frac{\partial^2{\bf H}}{\partial q^i\partial q^j} + \frac{\partial q^i}{\partial u}\frac{\partial p_j}{\partial v}\frac{\partial^2{\bf H}}{\partial q^i\partial p_j}
\nonumber\\
& & - \frac{\partial p_i}{\partial v}\frac{\partial q^j}{\partial u}\frac{\partial^2{\bf H}}{\partial p_i\partial q^j} - \frac{\partial p_i}{\partial v}\frac{\partial p_j}{\partial u}\frac{\partial^2{\bf H}}{\partial p_i\partial p_j}\Biggr)\;,
\end{eqnarray}
or, reorganizing terms,
\begin{eqnarray}
\frac{dJ_k}{dt} &=& \frac{1}{2\pi}\int_{{\cal S}_k} du\,dv\Biggl[\frac{\partial^2{\bf H}}{\partial p_i\partial q^j}
\left(\frac{\partial p_i}{\partial u}\frac{\partial q^j}{\partial v} - \frac{\partial q^i}{\partial v}\frac{\partial p_j}{\partial u}\right)
\nonumber\\
& & + \frac{\partial^2{\bf H}}{\partial p_i\partial p_j}
\left(\frac{\partial p_i}{\partial u}\frac{\partial p_j}{\partial v} - \frac{\partial p_i}{\partial v}\frac{\partial p_j}{\partial u}\right)
\nonumber\\
& & + \frac{\partial^2{\bf H}}{\partial q^i\partial q^j}
\left(\frac{\partial q^i}{\partial u}\frac{\partial q^j}{\partial v} - \frac{\partial q^i}{\partial v}\frac{\partial q^j}{\partial u}\right)
\nonumber\\
& & + \frac{\partial^2{\bf H}}{\partial q^i\partial p_j}
\left(\frac{\partial q^i}{\partial u}\frac{\partial p_j}{\partial v} - \frac{\partial p_i}{\partial v}\frac{\partial q^j}{\partial u}\right)\Biggr]\;.
\label{eq:dJkdt2}
\end{eqnarray}
The indices $i$ and $j$ are both dummy indices; we are using the summation convention, so there is an implied sum from $1$ to $N$ over both of these indices.  With this in mind, we see we see that every term in parentheses in Eq.\ (\ref{eq:dJkdt2}) sums to zero.  We thus find
\begin{equation}
\frac{dJ_k}{dt} = 0\;,
\end{equation}
proving adiabatic invariance.

As emphasized above, a crucial assumption in this analysis is that the dynamical system passes through every allowed point in its range as the Hamiltonian ${\bf H}$ evolves from one integrable configuration to another.  This assumption guarantees that the integrals defining $dJ_k/dt$ are well defined along the evolving sequence.

What if this assumption is not met?  In general, there is no simple way to handle this circumstance.  In the language of the problem that motivates this paper, one would need to solve for orbits in a rapidly evolving spacetime, a problem for which almost certainly there is no closed-form solution.  However, in the limit in which the underlying spacetime or potential changes so rapidly as to be practically instantaneous, a simple solution emerges: the system changes so rapidly that the canonical coordinates and momenta are unchanged during the transition: $({\bf p},{\bf q})^{\rm before} = ({\bf p},{\bf q})^{\rm after}$.

At least for black hole orbits, it is difficult to imagine a circumstance in which the spacetime would evolve so rapidly that the condition just described is relevant.  The slowly evolving case for which adiabatic invariance pertains is far more likely to be of astrophysical use.

\section{Computing Kerr orbit actions and their derivatives}
\label{app:actions}

To compute the actions $J_{r,\theta,\phi}$ for Kerr black hole orbits and the derivatives of the actions which are used in Sec.\ {\ref{sec:kerr_generic}}, we use the following computational recipe.  First, select the orbit's geometrical parameters $p$, $e$, $I$.  These parameters allow us to remap the radial motion to an anomaly angle $\psi$, and to remap the polar motion to an anomaly angle $\chi$:
\begin{eqnarray}
r &=& \frac{p}{1 + e\cos\psi}\;,
\label{eq:psidef}\\
\cos\theta &=& \sin I \cos\chi\;.
\label{eq:chidef}
\end{eqnarray}
The angle $\psi$ accumulates secularly as $r$ oscillates from $r_{\rm min} = p/(1+e)$ to $r_{\rm max} = p/(1 - e)$ and back; likewise, $\chi$ accumulates secularly as $\theta$ oscillates from $\theta_{\rm min}$ to $\theta_{\rm max}$ and back.  Neither $\psi$ nor $\chi$ exhibits pathologies associated with turning points in the motion (points when the coordinate velocity passes through zero and reverses).

Using formulas in Ref.\ {\cite{schmidt}}, we next compute the orbit's conserved integrals $E$, $L_z$, and $Q$.  Note that Schmidt uses the parameter $\theta_{\rm min}$ rather than $I$.  Since the definitions (\ref{eq:thetaminmax_pro}) and (\ref{eq:thetaminmax_ret}) apply to the Kerr black hole case, it is simple to convert (noting that $I \le \pi/2$ describes $L_z$ positive, and $I > \pi/2$ is for $L_z$ negative).  It is then straightforward to evaluate the action integrals:
\begin{eqnarray}
J_r &\equiv& \frac{1}{2\pi}\oint p_r\,dr = \frac{\mu}{\pi}\int_{r_{\rm min}}^{r_{\rm max}} \frac{\sqrt{R}}{\Delta}dr
\nonumber\\
&=& \frac{\mu}{\pi}\int_0^\pi \frac{\sqrt{R}}{\Delta}\frac{dr}{d\psi}\,d\psi\;,
\label{eq:Jrdef}\\
J_\theta &\equiv& \frac{1}{2\pi}\oint p_\theta\,d\theta = \frac{\mu}{\pi}\int_{\theta_{\rm min}}^{\theta_{\rm max}}\sqrt{\Theta}\,d\theta
\nonumber\\
&=& -\frac{\mu}{\pi}\int_0^\pi\sqrt{\frac{\Theta}{1 - \cos^2\theta}}\frac{d\cos\theta}{d\chi}\,d\chi\;,
\label{eq:Jthdef}\\
J_\phi &\equiv& \frac{1}{2\pi}\oint p_\phi\,d\phi\nonumber\\
&=& L_z\;.
\label{eq:Jphidef}
\end{eqnarray}
In these equations, the functions $\Delta$ and $R$ are functions only of the coordinate $r$,
\begin{eqnarray}
R &=& \left[E(r^2 + a^2M^2) - a M L_z\right]^2
\nonumber\\
& &\qquad - \Delta\left[r^2 + (L_z - a M E)^2 + Q\right]\;,
\label{eq:Rdef}\\
\Delta &=& r^2 - 2Mr + a^2M^2\;,
\label{eq:Deltadef}
\end{eqnarray}
and $\Theta$ is a function only of the coordinate $\theta$,
\begin{equation}
\Theta = Q - \cos^2\theta\,a^2M^2(\mu^2 - E^2) + \cot^2\theta\,L_z^2\;.
\label{eq:Thetadef}
\end{equation}
Thanks to our use of $a = S/M^2$, extra factors of $M$ may seem to be in these equations compared to their appearance in other literature.  This form is particularly useful for computing derivatives of the actions with respect to $S$ and $M$.

Using Eqs.\ (\ref{eq:Jrdef})--(\ref{eq:Jphidef}), it is a straightforward exercise to compute derivatives of the actions with respect to any of $p$, $e$, $I$, $M$, $S$, and $\mu$.  A {\it Mathematica} notebook which performs these calculations is freely available to any interested reader.

\section{Flux of radiation for circular, equatorial inspiral}
\label{app:flux}

We use the following analytic expansions for the fluxes of energy carried by gravitational waves from circular and equatorial orbits of Kerr black holes.  Note that many additional terms in these expansions are known (see, for example, Ref.\ {\cite{fujita2015}} for a recent summary and discussion); the truncated expressions given below are sufficient for the present analysis.

To begin, define $x \equiv (M\Omega)^{1/3}$.  The flux to infinity is then given by
\begin{eqnarray}
\dot E^\infty &=& -\frac{32}{5}\left(\frac{\mu}{M}\right)^2x^{10}\times
\nonumber\\
& &\left(1 + I_2 x^2 + I_3 x^3 + I_4 x^4 + I_5 x^5\right)\;,
\end{eqnarray}
where
\begin{eqnarray}
I_2 &=& -\frac{1247}{336}\;,
\\
I_3 &=& 4\pi - \frac{11a}{4}\;,
\\
I_4 &=& -\frac{44711}{9072} + \frac{33a^2}{16}\;,
\\
I_5 &=& -\frac{8191\pi}{672} - \frac{59a}{16}\;.
\end{eqnarray}
The flux down the event horizon is given by
\begin{eqnarray}
\dot E^{\rm H} &=& -\frac{32}{5}\left(\frac{\mu}{M}\right)^2x^{15}\times
\nonumber\\
& &\left(H_0 + H_2 x^2 + H_3 x^3 + H_4 x^4 + H_5 x^5\right)\;,
\end{eqnarray}
where
\begin{eqnarray}
H_0 &=& -\frac{a}{4} - \frac{3a^3}{4}\;,
\\
H_2 &=& -a - \frac{33a^3}{16}\;,
\\
H_3 &=& \frac{1+\kappa}{2} + a\left[2B_2(1 + 6a^2) + \frac{13a}{2}\kappa\right.
\nonumber\\
& &\left. + \frac{35a}{6} - \frac{a^3}{4} + 3\kappa a^3\right]\;,
\\
H_4 &=& -a\left(\frac{43}{7} - \frac{17a}{56} - \frac{4651a^2}{336}\right)\;,
\\
H_5 &=& 2(1 + \kappa) + a\left[B_1\left(1 - \frac{3a^2}{4}\right) + 6B_2(1 + 3a^2)\right.
\nonumber\\
& & \left. + \frac{433a}{24} + \frac{163\kappa a}{8} - \frac{95a^3}{24} + \frac{33\kappa a^3}{4}\right]\;.
\end{eqnarray}
In these expressions,
\begin{eqnarray}
\kappa &=& \sqrt{1 - a^2}\;,
\\
B_n &=& \frac{1}{2i}\left[\psi\left(3 + \frac{nia}{\kappa}\right) - \psi\left(3 - \frac{nia}{\kappa}\right)\right]\;,
\end{eqnarray}
where $\psi(z)$ is the polygamma function.


\begin{thebibliography}{99}

\bibitem{bt1987} J.\ Binney and S.\ Tremaine, {\it Galactic Dynamics} (Princeton University Press, Princeton, 1987).

\bibitem{carter1968} B.\ Carter, Phys.\ Rev.\ {\bf 174}, 1559 (1968).

\bibitem{bcp14} E.\ Barausse, V.\ Cardoso, and P.\ Pani, Phys.\ Rev.\ D {\bf 89}, 104059 (2014).

\bibitem{in2018} S.\ Isoyama and H.\ Nakano, Class.\ Quantum Grav.\ {\bf 35}, 024001 (2018).

\bibitem{bpt72} J.\ M.\ Bardeen, W.\ H.\ Press, and S.\ A.\ Teukolsky, Astrophys.\ J.\ {\bf 178}, 347 (1972).

\bibitem{tagoshi} H.\ Tagoshi, Prog.\ Theor.\ Phys.\ {\bf 93}, 307 (1995).

\bibitem{kennefick} D.\ Kennefick, Phys.\ Rev.\ D {\bf 58}, 064012 (1998).

\bibitem{schmidt} W.\ Schmidt, Class.\ Quantum Grav.\ {\bf 19}, 2743 (2002).

\bibitem{fujita2015} R.\ Fujita, Prog.\ Theor.\ Exp.\ Phys.\ {\bf 2015}, 33E01 (2015).

\bibitem{lob08} L.\ Lindblom, B.\ J.\ Owen, and D.\ A.\ Brown, Phys.\ Rev.\ D {\bf 78}, 124020.

\end{thebibliography}
\end{document}